# Dynamics of Abell 2218 from optical and near-IR imagery of arc(let)s and ROSAT/HRI X-ray map *

J.P. Kneib[1,2], Y. Mellier[1], R. Pelló[1,4], J. Miralda-Escudé[3], J.-F. Le Borgne[1], H. Böhringer[5], and J.-P. Picat[1]

[1] Observatoire Midi-Pyrénées, LAT URA-CNRS 285; 14 avenue Edouard Belin, F-31400 Toulouse, France
[2] Institute of Astronomy, Madingley Road, Cambridge CB3 0HA, UK
[3] Institute for Advanced Study, Olden Lane, Princeton NJ 08540, USA
[4] Laboratori d'Astrofísica del IEC, Diagonal 647, E-08028 Barcelona, Spain
[5] Max Planck Institut für Extraterrestrische Physik, D-8046 Garching, Germany

**Abstract.** A mass model of the rich cluster A2218 is presented based on optical and new near-infrared images of arc(let)s and the ROSAT/HRI X-ray map. A lensing model is proposed with two dark matter clumps, centered on the two bright galaxies around which the various arcs are observed. The number of free parameters of the model is reduced to four, namely two for the mass density profile of each clump. The centers, orientations and ellipticities of the mass distribution are fixed to those of the outer isophotes of the central galaxies. The model reproduces the four long arcs present in this cluster. Two of them, close to the most massive clump around the brightest galaxy, are required to be fold images (i.e., two merged images on a critical line), given their position angle with respect to the major axis. A third image for each of the two arcs is then predicted to be on the opposite side of the major axis. Both images are indeed found at the expected positions, with consistent magnitudes, colors and shapes. One of the arcs has a known redshift of 0.7, and the model predicts that the other arc should be at higher redshift, $z \approx 3$ (depending on the slope assumed for the density profile). Photometric data in UBrzJK' indicates that a redshift close to 3 is indeed probable for this arc. As in other cluster lenses (MS2137, Abell 370, Cl2236-04), the dark matter halo is found to be sharply peaked on the giant central galaxy.

The ROSAT/HRI X-ray map shows a strongly clumped image, poorly correlated with the cD light and the mass model on scales below $\sim 200 h_{50}^{-1}$kpc, but similar to the distribution of galaxies and the mass model when smoothed on this scale. This is expected if the cluster was formed by mergers and is not yet relaxed to a stationary equilibrium. The X-ray map is interpreted as due to gas falling along the major axis of the cluster towards a central shock, with a stream of shocked gas moving away from the center along the minor axis. The large degree of central substructure implied helps to account for the discrepancy previously found (in the two massive clusters A2218 & A1989) between the mass derived from the gravitational lenses and that derived from the X-ray profile and temperature distributions, in a model of hydrostatic equilibrium. Most of the discrepancy can probably be explained by bulk motions and acceleration of the gas, and projection effects enhancing the lensing power of the cluster.

## 1. Introduction

Gravitational lensing in clusters of galaxies has been used as a probe of the mass distribution (see Fort & Mellier 1994 for a review). When combined with optical, spectrophotometric, and X-ray data, the information given on the total mass by the lensing observations helps diagnose the dynamical state of the gas and galaxies in the cluster. Since the first discovery of lenses in clusters, it was apparent that the mass distribution cannot have a core radius larger than $\sim 50\, h_{50}^{-1}$ kpc (e.g., Grossman & Narayan 1989), and that the high central densities implied are dominated by dark matter, since the derived mass-to-light ratios are $M/L_B \sim 300 h_{50}^{-1}$ (see Table 2 in Fort & Mellier 1994).

The cluster dark matter must therefore be sharply peaked, near the cluster center around which the lensed images are seen. In many of the clusters where arcs are observed, the brightest cluster galaxy is, within the errors, exactly where this center is inferred to be from the lens modeling (as in A2218, A963, MS2137). These central galaxies are often cD galaxies, and in fact their extended light halo can be detected to a radius as large as the radial distance of the arcs ($\sim 100 \, h_{50}^{-1}$ kpc for A2218). The stars in these halos are therefore orbiting in the same dark matter potential well that causes the lensing and, if they are in dynamical equilibrium, their distribution must be related to the dark matter distribution. This suggests as a natural assumption that, if the dark matter has an ellipsoidal distribution, its major axis should be that of the stellar halo. The ellipticities could be different depending on the velocity anisotropy of the dark matter and the stars, but a first hypothesis is to assume that they are also identical.

These ideas were first applied to the clusters A370 by Kneib et al. (1993) and MS2137 by Mellier et al. (1993). In A370 two similarly bright galaxies are present and only a model assuming two clumps, with the ellipticities and centers of the two galaxy halos, is able to reproduce both the shape and the position of the long arc and of the two multiple images of another source. In MS2137 a tangential and a radial arc have been observed (Fort et al 1992). Mellier et al (1993), assuming a single clump for the mass, with the ellipticity and position of the largest galaxy, reproduce the positions and shapes of five images as arising from two background sources. This latter model, with only two free parameters for the lens (if the ellipticity, orientation and center are fixed), and four for the two source positions, can account for the ten coordinates of the five images, plus at least five observables for the shape of each image. The model is therefore highly predictive.

Here, we shall follow the same strategy to model the arcs in A2218, a cluster which has been extensively observed. This cluster is quite similar to MS2137 and A370 in the sense that it is dominated by a cD galaxy apparently centered on the cluster potential and also because it has at least two arcs with multiple image candidates. A2218 is one of the richest clusters in the Abell catalogue (Abell et al. 1989) and in the Butcher and Oemler photometric sample (Butcher et al. 1983; Butcher & Oemler 1984). Observations of the arcs and arclets in A2218 were first presented in Pelló et al. (1988, hereafter Paper I). Le Borgne et al. (1992, hereafter Paper II) have published a detailed photometric and spectroscopic survey of A2218. They have obtained a mean redshift of 0.175 and a velocity dispersion of $1370^{+160}_{-210} \, kms^{-1}$ from radial velocities of 50 cluster galaxies. This redshift is in good agreement with the first value measured by Kristian et al. (1978). The X-ray map of this cluster obtained by the Einstein Observatory (Boynton et al. 1982) shows a smooth profile and an almost circular symmetry on large scale. It is a strong X-ray emitter in the range 0.5-4.5 keV: $6.5 \times 10^{44} ergs \, s^{-1}$ (Perrenod & Henry 1981). The radio map of A2218 at 11.1 cm (Schallwich & Wielebinski 1978) shows an extended radio halo, with an intensity peak close to the center of the cluster. Andernach et al. (1988) have reviewed the observations of A2218 in the radio range, from 2.8 cm to 11.1 cm, and an optical identification of radio sources in this field can be found in Le Borgne & Vílchez-Gómez (1993). A2218 is also known to exhibit a Sunyaev-Zeldovich effect (Birkinshaw et al. 1981; Birkinshaw & Gull 1984; Partridge et al. 1987; Klein et al. 1991, Jones et al. 1993, Birkinshaw & Hughes 1994).

A more detailed study of the arcs and arclets in A2218 was done in Pelló et al. (1992) (hereafter Paper III). The identification of the various objects in this paper follows the numbering given in Papers II and III, which constitute the main optical database on this cluster. A total of 32 arc(let)s were identified in Paper III. These observations already suggested a model with two mass clumps centered on the two brightest galaxies, the clump on the cD being significantly more massive. The ellipticities and centers of these galaxies can be used as initial constraints on the mass distribution. The lensing model should account for the shapes of the various arcs and arclets, and predict the possible multiple images. There is also a luminous ring around galaxy #373, as discussed in paper III, which has been found by spectroscopy to be at the cluster redshift (Le Borgne et al. 1995, in preparation). In any case, it is difficult to interpret this as a gravitationally lensed image, given its almost perfect smoothness and circularity, and its proximity to the arc #359, which shows a very strong shear (Kassiola & Kovner 1993a).

The observational data used to constrain the model are briefly summarized in Section 2. Section 3 gives the constraints on the potential obtained from optical photometry of the cD halo, and presents new results from infrared photometry of the arcs and multiple image candidates. The lensing model is described in Section 4, together with the results about the main arcs. Section 5 is devoted to weak lensing as a probe of the matter distribution at large radius. Section 6 presents the X-ray observations and discusses the relation with the mass distribution given by the lensing model. A general discussion can be found in Section 7, together with the conclusions of this work. Throughout this paper, we assume $H_o = 50$ km sec$^{-1}$ Mpc$^{-1}$, $\Omega = 1$ and $\Lambda = 0$. The angular scale conversion at $z = 0.175$ is $1'' \equiv 3.83 h_{50}^{-1}$ kpc.

## 2. Summary of optical and near-infrared data and previous results

Optical photometric and spectroscopic data were obtained, respectively, at the 3.5 m telescope of the CAHA at Calar Alto (Almería, Spain) and at the 4.2 m WHT at the Observatorio Roque de los Muchachos (Canary Islands, Spain). All the details concerning observations and data reduction can be found in Paper II. Here we add the new near infrared photometry in J and K', obtained at the 3.6 m CFH telescope, of a few galaxies and arc(let)s which are interesting for the modeling. Results on infrared photometry of cluster galaxies will be given elsewhere (Le Borgne et al. 1995), together with details on data reduction and photometric calibrations. This J and K' photometry has been obtained with the Redeye camera in its wide field mode (0.5"/pixel, total field of view 2.1'×2.1'). The total magnitudes, surface brightness and colors of the most interesting objects in the context of the lens modeling are summarized in table I. Since we are mainly concerned with "photometric redshifts" and multiply imaged galaxies, we draw particular attention to galaxies with similar photometric properties. Individual objects will be discussed in Section 3 (observational constraints on the modeling).

In the present paper, we have mainly used the optical composite Brz images obtained in 1991 (with seeing $\sim 0.7'' - 0.8''$, and pixel size $0.253''$). 729 objects have been detected in the whole field. The central region of the cluster is dominated by a cD galaxy, but there is also a secondary bright clump concentrated around galaxies #244 and #259. The identification chart of all the arcs and arclets in A2218 is presented in Fig. 1 of Paper III, where arc(let)s are drawn by lines scaled to their lengths up to about $1\sigma$ of the sky noise in r. The geometric and photometric characteristics of the different arc(let)s (identification number and coordinates, magnitudes and colors, total length, position angles and photometric redshift obtained from Bgrz photometry) are summarized in Table 2 of Paper III.

Spectroscopic information was obtained for two of the long arcs around the brightest cD galaxy. One object (#359) has a spectroscopic redshift $z = 0.702$. The inferred spectral energy distribution of the object #384 (from a 5200-9000 Å spectrum and UBgrz photometry), prior to J and K' photometry, is compatible either with a foreground galaxy at $z \sim 0.1$ or with a very high redshift galaxy ($2.6 \leq z \leq 3.6$). Another system of two long arcs is seen close to galaxy #244, which is the second brightest galaxy, and which seems to have a faint extended stellar halo. In this second system, a bright region in arc #289 has a spectroscopic redshift of 1.034.

The configuration of these arcs strongly suggests a model with two mass clumps centered on these galaxies, the clump on the cD being significantly more massive, as shown by the distances of the arcs to galaxies. Another argument in favour of substructure in A2218 is the high peculiar velocity of the cD, $-657 kms^{-1}$ (see Malumuth 1992).

## 3. Constraints on the mass distribution obtained from optical and near-infrared observations

The main problem with the models of cluster lenses is the uniqueness of the solution. This problem is reduced to a large extent if optical observations of the galaxy halos and multiple images can be used to constrain the lensing model. Thus, we follow the same method as in Mellier et al. (1993) in the case of MS2137, and Kneib et al. (1993) in the case of A370. The photometry and the geometrical characteristics of the envelope of the cD galaxy are given to constrain the mass distribution, as described in the Introduction. We also give in Table I new results on the optical (Brz) and near-IR (JK') photometry of arcs (and multiple image candidates), together with the inferred photometric redshift. Objects are identified in Fig. 5a, according to the numbering given in Papers II and III. Photometric redshifts are calculated through colour-redshift diagrams obtained from Bruzual's code for the spectrophotometric evolution of galaxies (Bruzual & Charlot 1993), taking into account the transmission functions of the photometric system. As the spectral range of the observations has been increased to the near-IR, the permitted redshift intervals are more restrictive than those proposed in paper III. The mean surface brightness is obtained by considering all the pixels within the isophote corresponding to 1 $\sigma$ of the local sky noise. Colors are calculated from integrated isophotal magnitudes. The blue and red arc systems are given separately, according to the different cases considered in Section 4.

### 3.1. The envelope of the cD

The central cD galaxy has a single nucleus and an envelope well fitted by an $I(r) \propto r^{-1.57}$ law between $r = 2''.8$ ($11 h^{-1} kpc$) and $r = 25''.1$ ($96 h^{-1} kpc$), as expected for the brightness profile of such envelopes. Figure 1 shows the brightness profile of the cD in the r-band. Elliptical isophote fits are presented in Figure 2, as well as the best values of the model parameters (see Sect. 4). The ellipticity of the cD galaxy, defined as $e = (a - b)/(a + b)$, increases outwards. The innermost isophotes, at radius $r < 1''$ (about three times the seeing half-width), are circularized by the seeing, but the increase in ellipticity from about 0.10 to 0.3 at larger radius is real. The position angle of the major axis decreases monotonically with radius, and the total twist angle is about 8°, in rough agreement with the results by Porter et al. (1991). The position of the ellipse centroids with respect to the centroid of the first isophote also changes with radius, but it never exceeds $1''.5$.

The increase of ellipticity with radius is a general result found by Porter et al. (1991) for $\sim 50\%$ of their total sample of brightest E galaxies in 175 Abell clusters. According to these authors, there is observational evidence that the outer regions of giant elliptical galaxies in clus-ters are dynamically coupled to their parent clusters. The mass distribution deduced from lensing gives an independent and complementary approach to cluster dynamics at small scales.

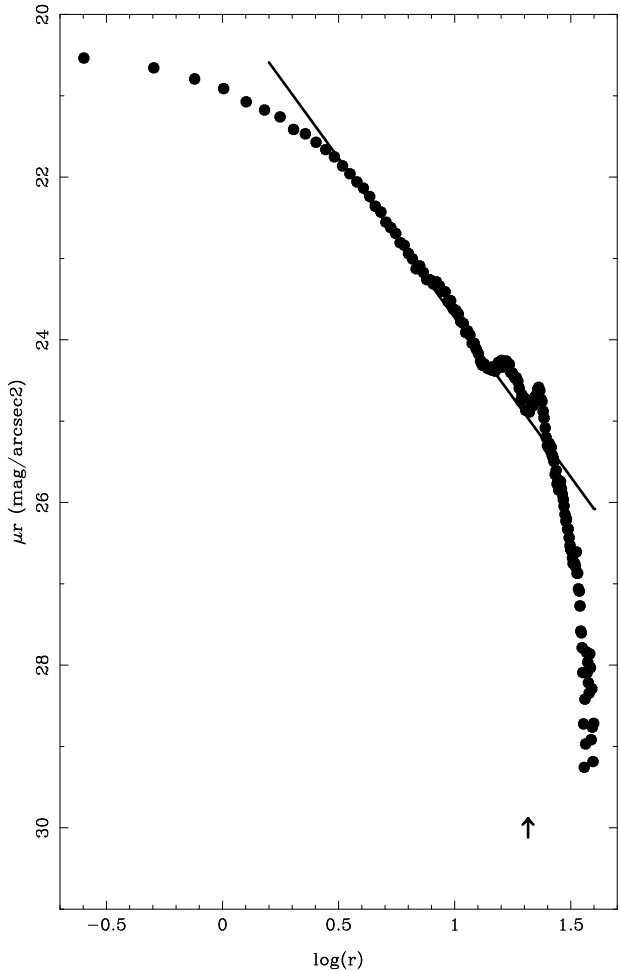

**Fig. 1.** r-band surface brightness profile of the central cD galaxy and its fit by an $I(r) \propto r^{-1.57}$ law. The plot displays $\log r = \log\sqrt{ab}$ in abscissa, where a and b are, respectively, the ellipse major and minor axes in arcsec. The position of the $z = 0.702$ arc is indicated by an arrow

We also looked carefully for faint, radially elongated images within the cD itself. Such images are expected if a background galaxy is on the radial caustic; they provide strong constraints on the radial profile of the lens within the region of the long arcs, and are always accompanied by a counter-image outside the tangential critical line. Only one such radial image has been discovered so far, in MS2137 (Fort et al. 1992). Because of the strong contamination by the cD light, we subtract from the B frame $\alpha$ times the r frame (instead of removing an analytical profile of the cD), choosing $\alpha$ to remove as completely as possible the inner part of the cD, where radial arcs could be present. This technique appeared to be very efficient in detecting blue or red faint galaxies underlying bright objects (Fort et al 1988; Kassiola et al. 1992, 1994). Indeed, the detection criteria require that some contiguous pixels are above the background, *all* having the same colour different from the cD colour. The capability to detect *by eye* any object corresponding to these criteria is high, although the magnitude estimate remains poor.

We cannot detect any radially elongated object in this region but the central part of the cD is much noisier than the sky signal. Besides, a residual $(I_B - \alpha I_r)$ remains in the cD envelope after subtraction because of a color gradient. So we estimate the detection limit as follows: to be detectable, any object within the cD should have at least 4 contiguous pixels one $\sigma$ above (if it is a red object), or below (if it is a blue object) the local background. This condition takes into account the pixel size ($0''.253$) and the seeing of the images ($0''.7$ to $0''.8$). An object is detected on this $B - \alpha r$ frame if its intensity in counts per pixel, $I$, is

$$I \geq \frac{\sqrt{(I_B + \alpha I_r)_{sky+cD}}}{2} + (I_B - \alpha I_r) \ .$$

If a detection is made, it is much easier to go back to initial B and r frames and compute the magnitude from the position and the shape inferred from the $I_B - \alpha I_r$ image. Using this technique, we conclude that no radial arc underlies the cD, which is longer than 0.5 arcsec, (2 pixels) and with a surface brightness higher than $B = 24.1 mag/arcsec^2$ and $r = 24.0 mag/arcsec^2$.

### 3.2. The red arc system around the cD : the arc #359 and its multiple image candidates

The spectrum of the arc #359 is identified as a non-evolved E/Sa galaxy at a redshift of z = 0.702 (see Paper III), thanks to several absorption features, such as the H and K lines, the discontinuity at 4000 Å and the CN bands. The spectroscopic redshift is fully consistent with the photometric UBgrzJK' redshift. It is worth noting that it is one of the reddest objects in the field, as well as one of the reddest arcs detected up to now. Figure 3 shows the spectral energy distribution of this object given by photometry, from U to K', as well as its fit by 2 synthetic spectra. The spectral energy distribution is consistent with a lack of star formation in the system for at least 2-3 Gyr. This is the only way to produce the observed break in the UBgr continuum. Nevertheless, the flux observed in the near IR indicates that the age of the last star formation burst is not older than 5 Gyr (otherwise, the flux in the zJK' bands should have been higher).

Among the 729 objects detected in the cluster, only 51 belong to the red interval $2.0 \leq B - r \leq 4.0$ and $0.5 \leq r - z \leq 1.5$, and only 29 of them have surface brightness of the same order as the arc #359 ($23.5 \leq \mu_r \leq 25.5$, $24.5 \leq \mu_B \leq 27.0$). The arc #359 is at $20''.7$ from the centroid of the first isophote of the cD. Within a radial distance of $40''.0$, only objects #389, #328 and #337 have colors and surface brightness estimates compatible with the arc #359 and can be considered as candidates for multiple images.

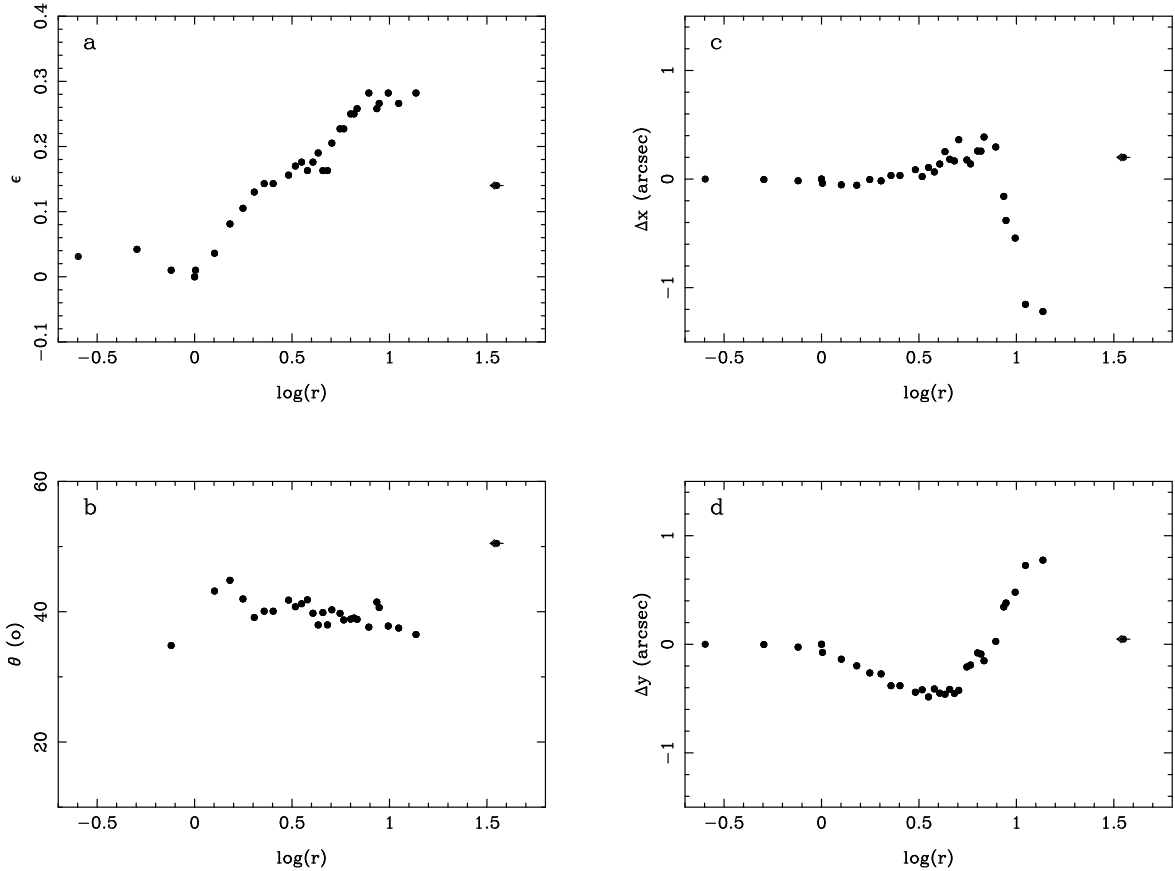

**Fig. 2.** cD isophote fits obtained in the r-band compared to the best model found for the mass distribution; the best model parameters are indicated by arrows: a) ellipticity (defined as $\epsilon = (a-b)/(a+b)$, where a and b are respectively the major and minor axes); b) position angle of the major-axis with respect to the N-S direction, increasing eastwards from S; c and d) position of the ellipse centroids relative to the centroid of the first isophote. As in Fig. 1, all the plots display $\log r = \log \sqrt{ab}$ in abscissa

### 3.3. The blue arc system around the cD : the arc #384 and its multiple image candidate

The most striking object in the northern region is #384. This object shows a break and a small curvature radius of about 13 arcsec. It is extremely blue compared to the cluster galaxies (see Table 1 and Paper III). Arc #384 is at $22''\!.1$ from the center of the cD. Compared to other objects detected in the whole field, there are 207 blue objects within the interval $0.5 \leq B-r \leq 1.5$ and $-0.5 \leq r-z \leq 0.5$, and only 142 with a surface brightness of the same order as this arc ($24.5 \leq \mu_r \leq 26.0$, $25.0 \leq \mu_B \leq 27.0$). This number reduces to 24 when the sample is limited to objects within a radial distance of $40''$.

A fit by a synthetic spectrum of the observed spectral energy distribution of object #384 is given in Figure 4. Although the optical spectrum was too noisy to determine the redshift, the continuum is in good agreement with the photometric data in the corresponding wavelength range. Two possibilities were considered in Paper III to account for the break between U and B: the low redshift hypothesis ($z \sim 0.1$), when it is identified with the 4000 Å break, and the high redshift hypothesis ($z \sim 3$), when it is the Lyman break. In the low-redshift case, the flux observed from U to z bands requires an active star forming system, such as an Im galaxy, but the flux detected in J and K' should be lower than observed, so this hypothesis is excluded. In the high-redshift case, the observed spectrum is compatible with a burst of star formation, probably no older than 1 to 1.1 Gyr. The Ly break between U and B requires a redshift higher than 2.6 and the absence of strong emission lines

**Table 1.** Photometric properties of image candidates around the cD. Internal errors are given below each photometric data

| Object | $\mu_B$ mag/ arcsec$^2$ | $\mu_r$ mag/ arcsec$^2$ | $\mu_z$ mag/ arcsec$^2$ | $\mu_J$ mag/ arcsec$^2$ | $\mu_{K'}$ mag/ arcsec$^2$ | B-r | r-z | r-J | J-K' | Photometric Redshift |
|---|---|---|---|---|---|---|---|---|---|---|
| #328 | 26.62 (0.20) | 24.77 (0.10) | 23.72 (0.10) | 22.14 (0.10) | 20.87 (0.10) | 3.59 (0.45) | 0.83 (0.15) | 2.82 (0.15) | 1.33 (0.25) | 0.65 - 1.13 |
| #337 | 25.19 (0.35) | 23.60 (0.10) | 22.79 (0.15) | 20.94 (0.25) | 19.65 (0.10) | 2.59 (0.50) | 0.74 (0.40) | 3.69 (0.50) | 1.68 (0.35) | 0.75 - 1.38 |
| #359 | 26.36 (0.15) | 24.10 (0.15) | 23.32 (0.15) | 21.41 (0.20) | 19.95 (0.10) | 3.11 (0.25) | 0.80 (0.35) | 3.19 (0.50) | 1.07 (0.50) | 0.63 - 1.20 |
| #389 | 25.82 (0.10) | 24.11 (0.10) | 23.41 (0.10) | 21.57 (0.10) | 20.18 (0.10) | 2.49 (0.35) | 0.93 (0.35) | 3.65 (0.30) | 1.25 (0.20) | 0.75 - 1.38 |
| #384 | 25.82 (0.20) | 24.89 (0.10) | 24.53 (0.10) | 22.71 (0.10) | 20.93 (0.10) | 0.83 (0.10) | 0.00 (0.30) | 1.56 (0.25) | 1.99 (0.20) | 2.60 - 3.30 |
| #468 | 25.89 (0.10) | 25.08 (0.10) | 24.78 (0.25) | 22.35 (0.60) | 21.13 (0.10) | 1.18 (0.15) | -0.12 (0.30) | 2.19 (0.15) | 1.73 (0.15) | 1.98 - 2.80 |

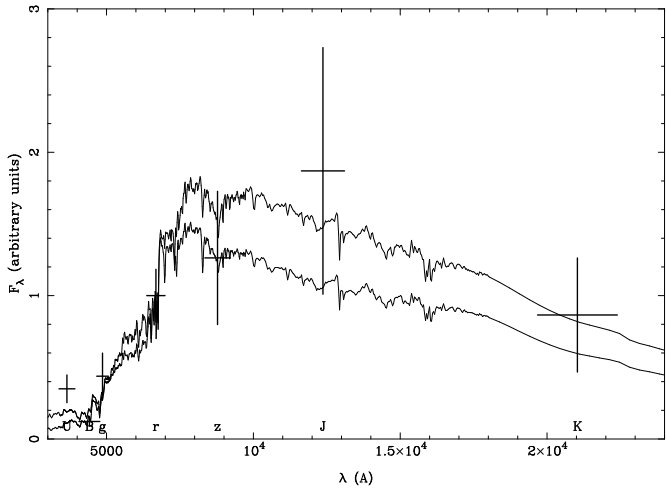

**Fig. 3.** Plots of the photometric points in the UBgrzJK' bands of the arc #359 with two synthetic spectra redshifted at z=0.702. The flux in the U band is an upper limit because the object is not detected in this band. The two synthetic spectra correspond to bursts of star formation aged 2 and 3 Gyrs, respectively

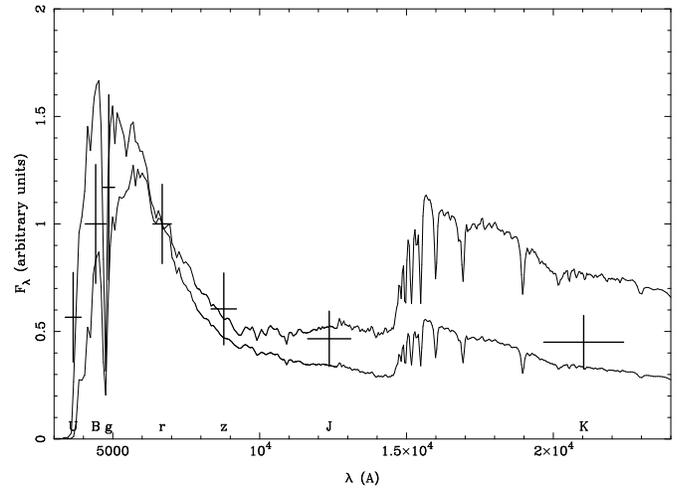

**Fig. 4.** Plots of the photometric points in the UBgrzJK' bands of the arc #384 with two synthetic spectra redshifted at z=2.9. The two synthetic spectra correspond to bursts of star formation aged 1.01 and a 1.14 Gyrs. The compatibility between photometric data and these redshifted spectra is remarkable

(Ly$\alpha$) in the 5200-9000 Å range implies a redshift lower than 3.3.

### 3.4. The arcs system around galaxy #244

Two large arcs #289 and #730 are seen around galaxy #244. The redshift of #289 is 1.034 (paper III), and no spectroscopic redshift exists for #730. However their colors differ and therefore are not multiple images of the same source. #289 looks patchy, and does not seem to continue in the halo of #244. The absence of interesting features on these arcs make them difficult to use as contraints for the modeling. However, since the curvature radius of #289 is opposed to the curvature of the arc #730, and also opposed to what is expected from the cD potential alone, it is inferred that a second massive clump exists which should be centered close to #244.

## 4. Modeling the central mass distribution

### 4.1. Modeling techniques and assumptions

The modeling is based on a standard $\chi^2$ minimization (Mellier et al. 1993, Kneib et al. 1993) of the difference between the source parameters of each multiple image. The fit includes all the observational constraints derived from the optical data of the cluster galaxies, the shape parameters of the lensed images and, whenever possible, the location of breaks at the position of merging images. The advantages of this technique are speed and simplicity, but minimization in the image plane should be more accurate.

We assumed a bimodal mass distribution, with a dominant clump centered on the most extended isophote of the cD, and the other centered on the bright galaxy #244. We used the pseudo-isothermal-elliptical-mass-distribution (PIEMD, Kassiola & Kovner 1993b) which is an exact elliptical mass distribution, and can be used for large ellipticities. The projected mass profile $\Sigma$ is:

$$\kappa = \frac{\Sigma}{\Sigma_{crit}} = \frac{E}{2\sqrt{r_c^2 + \rho^2}},$$

$$\text{with} \quad \rho^2 = \frac{x^2}{(1+e)^2} + \frac{y^2}{(1-e)^2},$$

where $\kappa$ is the convergence, $\Sigma_{crit} = \frac{c^2}{4\pi G} \frac{D_{OS}}{D_{OL} D_{LS}}$ is the usual critical mass density, $e = (a-b)/(a+b)$ is the ellipticity of the mass distribution, $r_c$ the core radius and in the limit $e \to 0$, $r_c = r_{1/2}/\sqrt{3}$ where $r_{1/2}$ is the half-maximum radius of the projected mass density. The deflection angles and amplification matrices are calculated using the formalism of Bourassa & Kantowski (1975). $E/r_c$ is the Einstein radius for $r_c \to 0$ and $e \to 0$. At large radii $r \to \infty$ we have:

$$E/r_c = 4\pi \frac{\sigma_\infty^2}{c^2} \frac{D_{LS}}{D_{OS}}.$$

The velocity dispersion (assumed isotropic) decreases towards the center. The central velocity dispersion is: $\sigma_0 \approx \sigma_\infty/1.45$. For the same central density and the same half-maximum radius as the density profile used by Mellier et al (1993) and Kneib et al (1993), this profile has more mass at large radius, although the slope at large radius is the same (see comparison of PIEMD & PIEP in Kassiola & Kovner 1993b). It is therefore difficult to make a precise comparison between the two types of mass profile.

The orientation and ellipticity of the two PIEMD potentials are those of the external isophotes of galaxies. During the modeling, all the geometrical parameters of the two clumps are kept free but within bounds deduced from the observational errors, and only the core radius and the velocity dispersion are left completely free. However, the curvature of the two arcs give an upper constraint on the core radius $r_c \lesssim 20"$, and since the redshift of #359 is known, the radius of the tangential critical line (which depends on $\sigma_0^2$ and $r_c$) is also strongly constrained. Moreover, because no radial arc was found in the cD halo (Sect. 3.1) and no counter-arc candidates were found in the halo for the two fold arcs #359 and #384 (Sect. 3.2 and Sect. 3.3), the naked cusp configuration is the most likely. This configuration appears when the mass distribution has a strong ellipticity and a relatively large core radius. Because we do not know the redshift of #384, we keep constant its value during the first fit. Later, the redshift space of the source has been mapped out to find the best value.

According to section 3.2, different counter-image candidates could be associated with the red arc #359. So, in a first step, we apply the model only to explore the different possibilities and implications. In this first model, it is possible to reproduce the image #389 as a continuation of #359. In this case, a long fold arc is produced where #359 is only half of it, the rest being located behind the galaxy #373 and the ring, and object #389 is the visible end of the second half. This model produces also a third image which coincides with #328. In a second model, objects #337 and #359 are taken as two images of the same source. A cusp arc is produced, and the model predicts a third image, elongated in the direction given by #337 and #359, which could be associated to #389. Again, most of this image is lying underneath the galaxy #373. But, in this scenario, object #337 should have been more elongated than observed, unless the image is perturbed in a complicated way by the bright galaxy #341. A third model, with only objects #328 and #359 (two merging images), leads to a fold arc and it does not require any additional perturbing mass.

The first two models lead to an extension of object #389 underneath the galaxy #373 and the ring, which is invisible in the optical bands. Furthermore, the J and K' images definitely show that no extension of this object is underlying the blue ring around galaxy #373. Thus, any model which assumes that #337, #359 and #389 are images of the same source seems incompatible with the optical and near infrared data, except if we make use of the local lensing effect of galaxies #341 and/or #373. This model cannot be excluded, but it is more difficult to manage and it does not modify noticeably the absolute mass distribution inferred for the main clump. Therefore, we will no longer discuss this hypothesis. Object #328 is retained as the best image candidate, because it does not require any additional clump of mass, and it is mentioned hereafter as #359'.

As mentioned in Sect. 3.3, the blue arc #384 shows a break and is likely a fold configuration. The minimization procedure is applied to reconstruct simultaneously the multiple arc #359 and #359', assuming it corresponds to a fold configuration at redshift $z = 0.702$, and the double merging image #384. The small break in the r band is used to mark out the position of the critical line and implies, if the potential is centered on the cD, a redshift beyond $z = 2.5$ for this arc.

The clump #244 is constrained by the fact that #289 is assumed not to have any counter images, and by requiring that the shear orientation is given by the orientation of the arc #730.

### 4.2. Results

The best results of the modeling are given with error bars in Table 2 and in Fig. 5. The model can reconstruct simultaneously the images of the source at $z = 0.702$ and the blue arc #384. The reduced $\chi^2$ is small, showing that the fit is satisfactory. As in the case of MS2137 and Abell 370, the model successfully predicts the position and the shape of the third image of the arc #384, which is identified with object #468, and its redshift, $z = 3. \pm 0.2$. As we can see from Table 1, the colour and surface brightness of this image match perfectly those of the arc #384. This gives further support to our model and the assumptions we have made. In fact, the constraints on the redshift depend on the slope assumed for the projected mass density profile. If all other parameters are kept constant, and if the mass increases faster than the PIEMD profile, the redshift of the source could be lower than 3. Conversely, for a steeper mass profile its redshift could be higher. We then estimate the redshift range by doing a sequence of models of the arc #384 with various slopes for the mass profiles. We find that the redshift of the source should be $z = 3. \pm 0.6$ for "realistic" models, corresponding to a projected surface density $\Sigma \propto r^{-\alpha}$, with $0.7 \leq \alpha \leq 1.3$. This value is fully consistent with the photometric redshift given in Sect. 3.2.

The second clump is not constrained enough by the lensed images to give accurate values of the potential parameters. Nevertheless the shear map (Fig. 5a), which directly displays the expected orientation and shear intensity, is in good agreement with the orientation of the lensed images #289 and #730. High resolution images are needed to put stronger constraints on the mass distribution in this region.

The shear map of the overall field shows a remarkable compatibility with the arclets observed in the field (Paper III, Fig. 1 and 2 of Paper II). Note that the existence of a complete Einstein ring around the galaxy #373 is forbidden in view of these results.

The absence of a break in the fold image #359 demands that the central peak of the source is outside the caustic. A break is only found when the center of the source (with maximum surface brightness) is within the caustic, and is imaged in the fold arc. When the center is outside the caustic, we expect that the maximum surface brightness is lower in the fold arc (#359) than in the counter-image (#359'), but the resolution of our images is not enough to check this. The regularity of the surface brightness profile along the arc suggests that the source is an elliptical galaxy, in agreement with its spectrum.

The redshifts of the multiple image candidates of the arcs #359 and #384 are within reach of 4m-class telescopes. Their measurement should be the final test for the predictions of our model.

## 5. Weak lensing

The central arcs and possible multiple images constrain only the central part of the mass distribution. In Paper III, a total of 32 arclets were presented, which were selected as the objects with the largest axis ratios visible in the field. About half of them match the shear map in the central part (see Fig. 5), but at larger radius they usually show much stronger distortion than would be expected for an isothermal mass profile (see Fig. 6). This sample may therefore be highly contaminated in the outer part of the cluster by edge-on spirals or irregular galaxies in the foreground. Selection criteria is thus a critical point when trying to probe the distortion at larger radius, and only statistical analysis can lead to robust results.

Any galaxy lying behind a mass distribution will be distorted perpendicularly to the local gradient of the gravitational potential. At large radius from the cluster center, where the surface density is much less than the critical value, the distortion causes only a small change in the shape of the lensed images. This effect, known as "weak lensing", was first seen by Tyson, Valdes, & Wenk (1990) in the clusters A1689 and Cl1409+52, and more recently in other clusters (MS1224+20: Fahlman et al 1994, Cl1455+22 & Cl0016+16: Smail et al 1994, Cl0024+17: Bonnet et al 1994).

To measure the weak lensing effect in A2218, we select from the photometric catalog of Paper II a list of background galaxy candidates using photometric criteria. First, the permitted color range for galaxies at $z = 0.175$ is obtained: the spectrum of a single burst of star formation, 10 Gyr old, gives the reddest end of the interval whereas the bluest end is calculated from an Im-type spectrum. As the maximum photometric error expected is about 0.3 magnitudes, the permitted color ranges for cluster members are $0.37 \leq B - g \leq 1.22$, $-0.13 \leq g - r \leq 1.34$, $0.54 \leq B - r \leq 2.25$ and $-0.06 \leq r - z \leq 0.82$, which roughly correspond to the reddest end $+0.3$ magnitudes and the bluest end $-0.3$ magnitudes. When an object has all its colors within these permitted intervals, it is considered as a cluster member; otherwise, when there is at least one color outside the range, it is classified as a foreground or a background object candidate, and most objects outside the cluster are probably background galaxies (stars are excluded from the catalogue).

All cluster galaxies with known spectroscopic redshifts have been properly classified as cluster members, and most arclets in Paper III appear as non-member objects. This automatic procedure excludes from the catalogue of background galaxies all the objects with photometric redshifts compatible with $z = 0.175$. Nevertheless, it is evident that some objects will be improperly classified as cluster members when their permitted redshift intervals are wide, so in general we tend to lose background galaxies. Conversely, cluster members with atypical spectra could be misclassified as background galaxies, but this case should not be very common. Finally we might also be contaminated by low surface brightness foreground galaxies (Ferguson & McGaugh 1994).

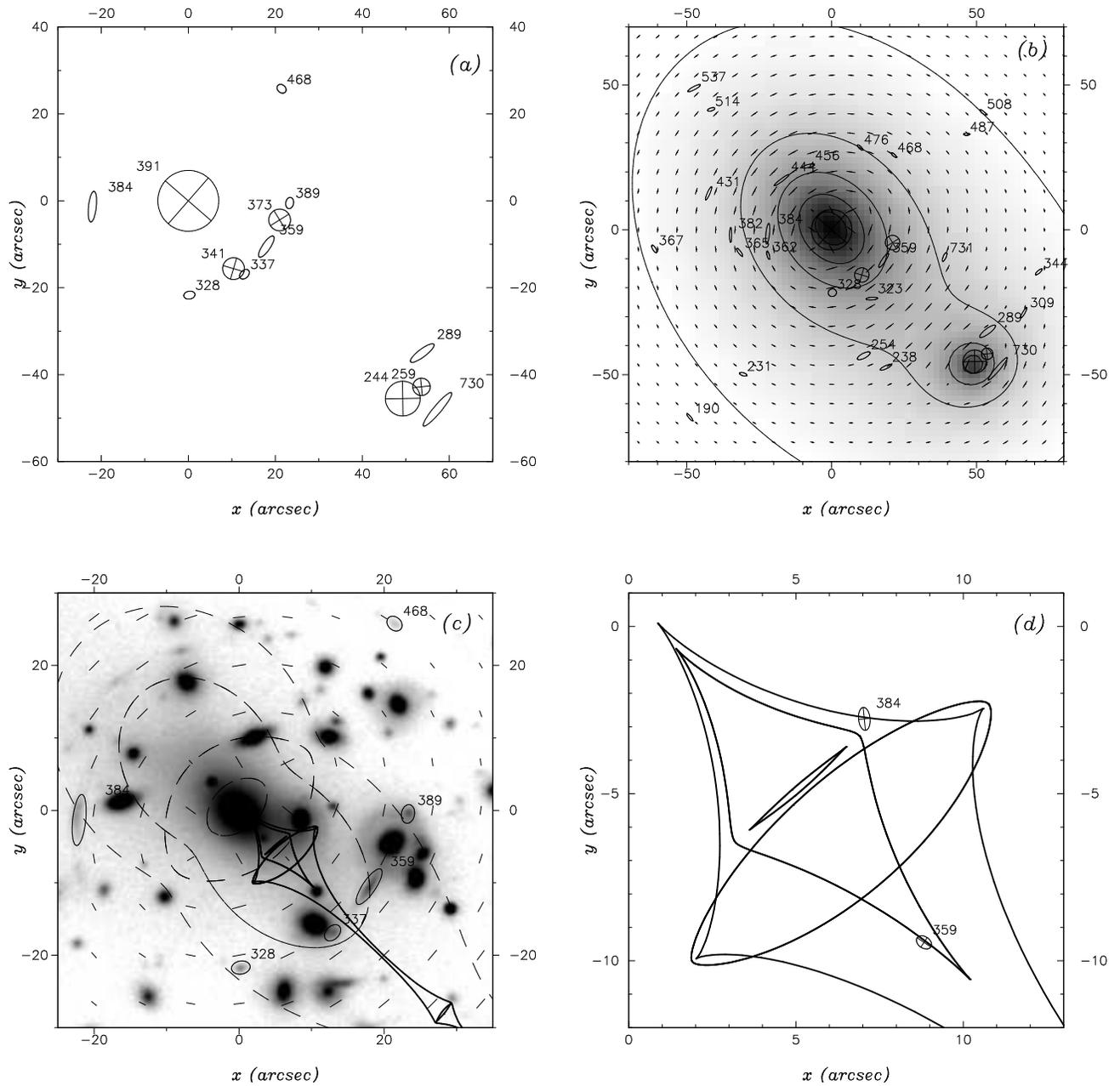

**Fig. 5.** Results of the modeling of A2218: a) Schematic diagram of the central part of A2218. Circles represent cluster galaxies, ellipses represent arc(let)s discussed in Section 3 and 4. b) Schematic diagram of the position of all the arclet candidates from Pelló et al. 1992, with the shear pattern and the iso-mass derived from the lens modeling. Note the shear field distortion around galaxies #244 and #259, which follows exactly the orientation and elongation observed in the lensed images. c) r-band image of the cluster, with the shear map and critical lines (for z=0.702 and z=2.8) superimposed. d) Schematic diagram of the source plane for the sources of arc #359 and #384.

**Table 2.** Parameters of the two projected potential found for the best model of the two arcs. The reduced $\chi^2 = 0.8$ gives a confidence level of 0.90. Above each quantity found by the fitting, we give the values of the same parameters of the galaxies associated with the center of the potentials.

|  | $x_c$ arcsec | $y_c$ arcsec | $e$ | a/b | $\theta$ degree | $r_c$ $h_{50}^{-1}\ kpc$ | $\sigma_\infty$ $km\,s^{-1}$ |
|---|---|---|---|---|---|---|---|
| cD halo | 0. | 0. | 0.145 | 1.34 | $139^\circ$ | - | - |
| model | -0.3 ±0.2 | 0.5 ±0.2 | 0.14 ±0.01 | 1.32 | $129.^\circ\ \pm 0.5^\circ$ | 45. ±2 | 1202. ±2 |
| #244 | 49.3 | -45.7 | 0.045 | 1.10 | $10.8^\circ$ | - | - |
| model | 48.9 ±0.5 | -46.6 ±0.5 | 0.05 ±0.03 | 1.11 | $18.0^\circ\ \pm 10^\circ$ | 19. ±5 | 649. ±10 |

For each of the candidates for background galaxies, we determine the axis ratio, $q$, and position angle of the major axis, $\theta$, through the first and second order moments of the 2D connected domain of the object (defined by all the contiguous pixels one $\sigma$ above the local sky background). We also measured the ellipticities by fitting the observed surface brightness within this domain to elliptical isophotes with exponential profiles, and found no substantial differences. The tangential component of the ellipticity is then defined as $\epsilon_t = (1-q^2)/(1+q^2) \cos(2\theta)$, where $\theta$ is measured from the tangential direction with respect to the central cD galaxy. This is equivalent to the quantity defined in Tyson et al. (1990) in terms of the second moments of the objects. For a circular source, $\epsilon_t$ is given in terms of the convergence and the shear by $2(1-\kappa)\gamma/[(1-\kappa)^2 + \gamma^2]$, which approaches $2\gamma$ in the weak lensing limit.

In Figure 6, we plot the quantity $\epsilon_t$ for each object, as a function of the distance $r$ from the cD galaxy. The black squares identify the 32 arclets given in Paper III, and the crosses correspond to all other objects selected as background galaxies. We have divided all the objects into three radial bins: $20'' < r < 40''$, $40'' < r < 80''$, and $r > 80''$. The thin solid lines give the average values of $\epsilon_t$ for all the objects, and the thick lines give the same average when the arclets from Paper III are excluded.

The value of the average "tangential alignment" measured is $\epsilon_t \simeq 0.07$, substantially less than expected in our model, if the background galaxies are reasonably distant from the cluster ($z > 0.5$). In fact, at redshift $z = 0.6$ ($z = 0.3$), the value of $\epsilon_t$ expected in our model would be $\sim 0.3$ ($\sim 0.17$) for most of the objects at an angular distance $r \sim 100''$. Part of this difference could be due to the seeing, the relatively low mean redshift of our galaxy sample, and contamination by cluster and foreground objects. However, we have argued that our color selection should leave us with few cluster objects, and the number of foreground objects is expected to be very small. The effects of seeing could play a large role, although our image is taken in good seeing conditions (0.7"). Therefore it should not be much worse than in most other cases where weak shear has been detected (Fahlman et al. 1994, Smail et al. 1994, Bonnet et al. 1994). They found that the measured shear is ≈70% the true value. However, some of these previous weak lensing studies were done in slightly better seeing condition with better sampling, 0.55" with 0.2"/pixel, instead of 0.7" with 0.253"/pixel. Our small detected shear could also be due to a different algorithm for measuring ellipticities compared to other work, but as we have mentioned we used two different algorithms which gave the same result. The weak shear techniques analyse the shape parameters of much fainter objects than ours and, consequently, they use more accurate algorithms to measure the geometry at very low flux levels, disregarding the photometric accuracy (Bonnet & Mellier 1995). Here the sample of no-cluster members has been selected through photometric criteria, which cannot be extended to the faintest objects of the field. This could explain the discrepancy between the present results and the ultra-deep weak lensing approaches.

The shear predicted by our model could also be too large, if the cluster surface density falls faster at small radius, and more slowly at large radius (since the shear is proportional to $\bar{\Sigma} - \Sigma$). This could be the case if the long arcs at $r \simeq 20''$ were produced by a chance alignment of two concentrated mass clumps, but the cluster mass distribution was more regular at the larger scale $r \sim 100''$.

In any case, it seems necessary that the consistency between the models for the highly magnified images and the weak lensing, as well as variations of the weak lensing from cluster to cluster, need to be understood before one can reliably measure masses of clusters of galaxies at large radius using weak lensing (Kochanek 1990, Miralda-Escudé 1991, Kaiser & Squires 1993).

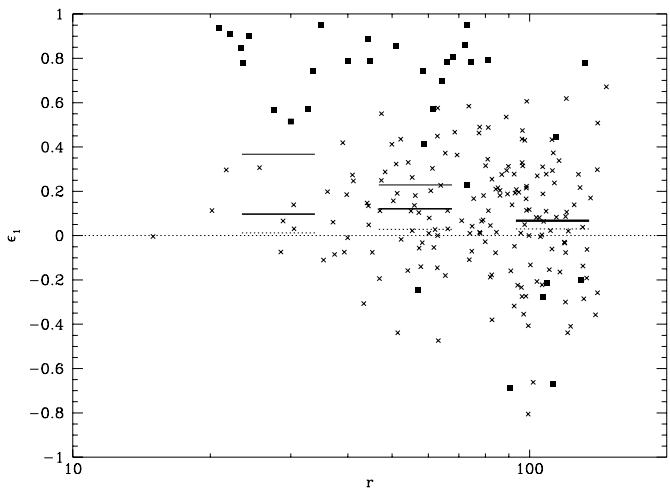

**Fig. 6.** Tangential component of the ellipticity (see definition in Sect. 5) as a function of the radial distance to the cD. Black squares identify the 32 arclets in Paper III, whereas crosses are the values for objects outside the cluster. The thin solid lines give the average values of $\epsilon_t$ for all the objects outside the cluster, the thick lines give the same average when the arclets from Paper III are excluded, and dashed lines give the average for the remaining sample of cluster members

## 6. Compatibility with the X-ray gas distribution.

### 6.1. The X-ray observations.

A2218 was observed with the ROSAT HRI (high resolution imager) on January 5-7, 1994 for 11,527 sec. In total about 1250±50 source counts were received from A2218 in the ROSAT window from about 0.1 to 2.4 keV. The nominal attitude accuracy of the ROSAT observatory pointing is about 6". In this exposure the stars HD 150098 and SAO 17153 were the only clearly detected stellar sources in the field of view of the HRI. These sources have been used to check and correct the data for positioning of the observational pointing. After a positional correction of about 3.5" to the north-east, the attitude accuracy is now better than 2".

Fig. 7 shows a contour map of the HRI X-ray image of A2218 superposed on the optical image. The X-ray image was smoothed by a Gaussian filter with a sigma of 8" (FWHM = 18.4"). The X-ray image shows a rather complex morphology. Instead of showing two X-ray maxima as one might have expected from the shape of the double elliptical model potential, only one maximum appears close to the position of the cD galaxy. At intermediate radii of 1' to 3', the X-ray contours are elongated with a position angle of about $135^\circ$ (from North to East) in approximately the same direction along which a chain of prominent bright galaxies are observed, including the cD and #244. In the central region the contours are elongated in approximately the east-west direction, almost perpendicularly to the elongation at intermediate radii. At larger radii the east-west direction is again pronounced in the contours.

**Fig. 7.** Smoothed ROSAT/HRI X-ray map of A2218 superimposed on the r-band CCD image. The X-ray contours give the sigma significance of the source counts above the background per filter area, the contour levels are $1.5\sigma$, $2\sigma$, and increase from thereon by $0.5\sigma$. The orientation and ellipticity of the X-ray contours roughly follows the cluster galaxy distribution on a large scale. But on a smaller scale, the orientation of the X-ray isophotes is perpendicular to the orientation of the cD light.

The maximum as well as the median of the spatial X-ray distribution is displaced somewhat to the west compared to the bright chain of galaxies by about 25". The emission as traced by the contours shown in Fig. 7 covers a region with a diameter of about $1h_{50}^{-1}$ Mpc.

### 6.2. On the dynamical stage of the X-ray gas.

The fact that the small scale X-ray gas distribution (see Fig. 9) does not resemble the image expected for gas in equilibrium in the gravitational potential of the lens model – as was observed in the case of A370, with two X-ray maxima centered on the two potential minima of the model (Mellier et al. 1994) – leads us to propose that the gas in this cluster could be severely disturbed by an ongoing cluster merger. N-body/hydrodynamical simulations of the hierarchical growth of clusters by mergers of smaller units by Schindler and Müller (1993) show very similar configurations. In particular, their model showing the merging of a chain of three major subclumps along one direction, which looks very similar to the matter distribution in A2218 indicated by the chain of bright galaxies. The sequence of model configurations in Figure 5d of Schindler and Müller (1993) shows X-rays features which are relevant for our observations. At intermediate distance, we recognize an elongation of the X-ray isophotes near the merger (following the distribution of the infalling matter) and also some elongation perpendicular to that, due to a shock wave and the subsequent gas expansion. The shock wave starts with a lenticular shape with the larger extension perpendicular to the collision axis. The resultant X-ray contours resemble the ROSAT image of A2218 very closely.

In this scenario, we can study the possible role of the galaxy #244. If it was the central dominant galaxy of a pre-existing subcluster, the galaxy should have preceded the gas during the final infall. This then seems to imply that the galaxy has already gone once through the new cluster center, and then a large part of the gas would have been stripped and would be found in a shock in the central cluster region. This is what is visible in the configuration of Fig. 5d of the model by Schindler and Müller: the second largest clump has just passed through the center and is now found in the south-east, stripped of its gas halo.

It can be concluded from this discussion, even without going into details, that the X-ray map obtained for A2218 can be very well explained by an ongoing cluster merger. Large deviations from hydrostatic equilibrium can occur in this case, and one can expect that the gas does not trace the gravitational potential.

## 7. Discussion and conclusions

The observational constraints provided by the multi-band optical and near-IR observations allowed us to build up a predictive model of the projected potential in A2218, and we can use it to discuss the dynamics of A2218.

This model can be compared with the two previous predictive models for cluster lenses obtained for MS2137-23 and A370. These three clusters are dominated by bright giant ellipticals or cD galaxies. In all cases the models support the same hypothesis: the external isophotes of the brightest ellipticals trace the orientation and the ellipticity of the projected potential, and the cluster center is very close to the centroid of the brightest isophote of the galaxy. This result indicates that the outer stellar halos of central cluster galaxies probe the dark matter distribution on a scale of $100h_{50}^{-1}$ kpc. At the same time, the stellar velocity dispersion of these halos must rise rapidly with radius, reaching values close to that of the cluster galaxies at the radius of the arcs. The central galaxy in A2218 is one of the brightest ones that have been observed in lensing clusters, and therefore it is a good candidate for measurement of the stellar velocity dispersion. The stellar central velocity dispersion gives a lower limit for the core radius (Miralda-Escudé 1994), for any given density profile. The lensing model implies an upper limit $r_c \leq 90h_{50}^{-1}kpc$, .

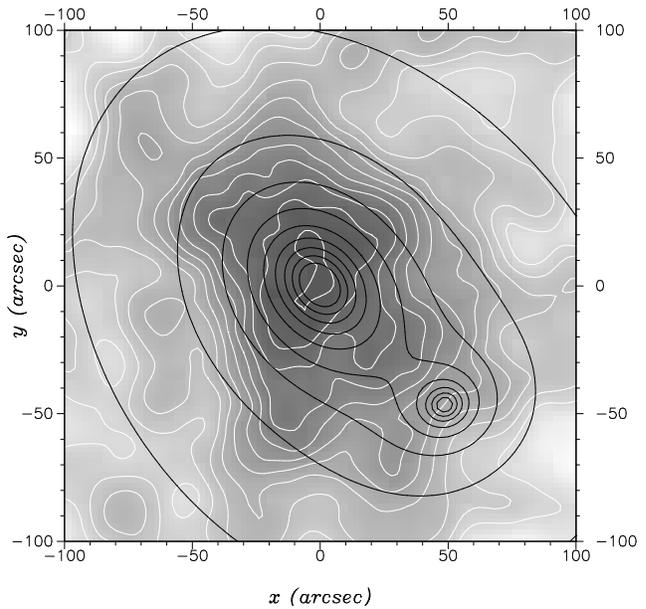

**Fig. 9.** Superposition maps of the HRI/ROSAT x-ray image and isocontours with the mass distribution model.

We have also measured the weak shear up to a radius $r = 130''$, the size of our image. We find that the shear at $r \sim 100''$ is significantly lower than expected from our model. Further observations and more careful analysis (including seeing correction) are needed to find the cause for this discrepancy.

The total mass found within the ellipse traced by the arc #384 ($r \approx 85kpc$) is $6.1\ 10^{13}M_\odot$ and gives a mass-to-light ratio $M/L_r = 80$. The most extended ellipse where the model is contrained by arcs reaches the second clump. At this distance from the cD, the ellipse contains $2.7\ 10^{14}M_\odot$ ($r \approx 256kpc$), and the mass-to-light ratio has increased up to $M/L_r = 180$. Beyond this region, the mass and the mass-to-light ratio continue to increase; but the model is not contrained anymore by the arcs and the mass profile is questionable. However, Fig. 8 shows a continuous increase of $M/L_r$ which rules out a simple relationship between mass and light within the innermost regions of A2218.

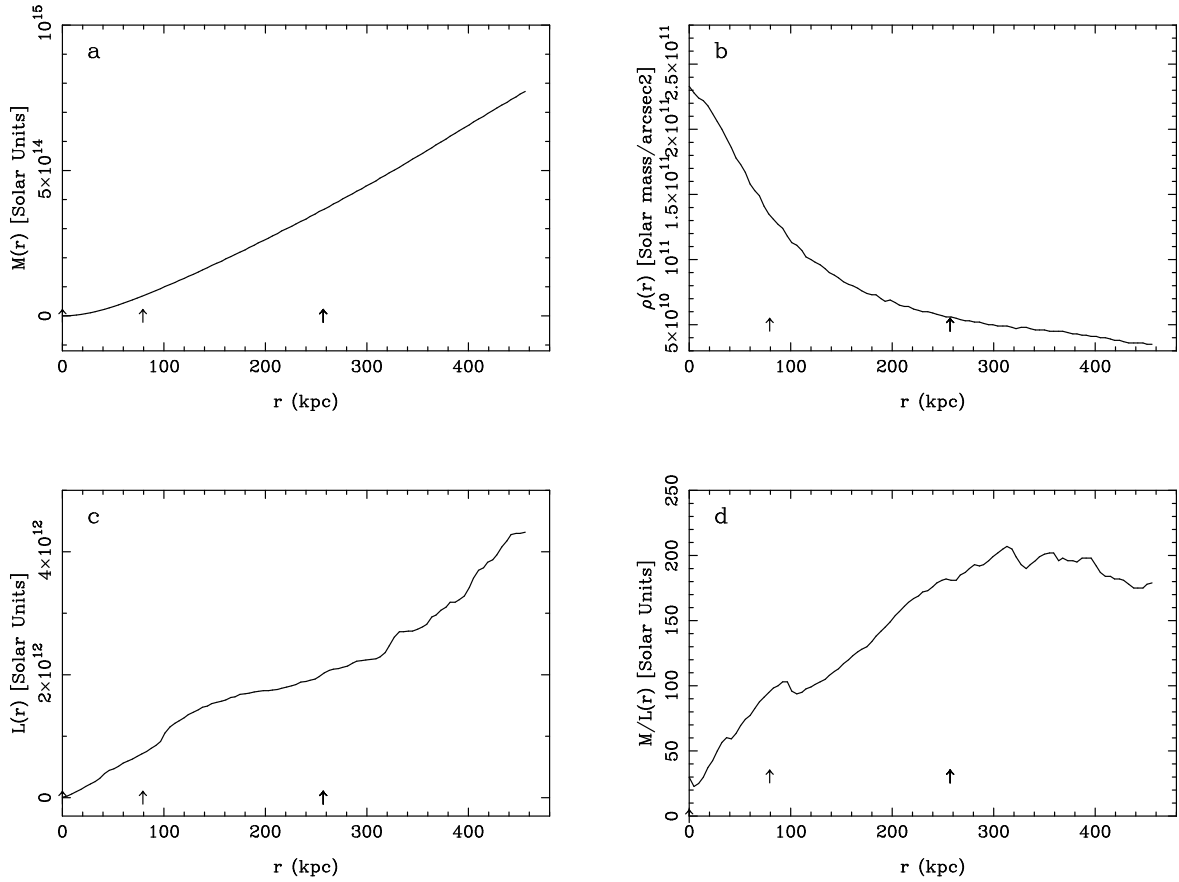

**Fig. 8.** Comparison of the total mass distribution with the total red light distribution: a) The total cumulated mass inferred from the model. Beyond 100" ($383h_{50}^{-1}$kpc), we do not have any constraints from giant arcs to extrapolate the increase of mass. At this distance we have about $4.5\ 10^{14}\ M_\odot$. b) Mass density profile. c) Cumulative total r-band luminosity. d) Variation of the mass-to-light ratio with radius, computed within elliptical annuli centered on the cD. It increases with radius up to a value of about $M/L_r \sim 200$ at $r \sim 400 h_{50}^{-1} kpc$

**Table 3.** Comparison between the 3 models computed with the same algorithm - Note that the mass profiles are not exactly the same -

| Parameter | A2218 cD clump | MS2137-23[1] cD clump | A370[2] #35/#20 | Cl2236-04[3] clump A |
|---|---|---|---|---|
| $r_{1/2}$ ($h_{50}^{-1}$kpc) | 82 | 35-55 | 80/95 | 52 |
| $\sigma^{obs}$ ($km s^{-1}$) | 1370 | 960* | 1350[†] | unknown |
| $\sigma_0^{model}$ ($km s^{-1}$) | 830 | 950-1200 | 845/820 | 610 |
| $\sigma_\infty^{model}$ ($km s^{-1}$) | 1202. | 1165-1530 | 1034/1004 | 750 |
| $(a/b)_{cD}/(a/b)_{model}$ | 0.98 | 0.99 | 0.98/0.95 | 1.0 |
| $\theta_{cD} - \theta_{model}$ (°) | 10. | 0. | 2./4. | 0. |
| $L_X$ ($10^{45} ergs\ s^{-1}$) | 0.7 | 0.7 | 0.8 | unknown |
| $z_{lens}$ | 0.175 | 0.313 | 0.374 | 0.56 |

[1] Mellier et al 1993, [2] Kneib et al 1993, [3] Kneib et al 1994, * Le Borgne et al 1995 *in preparation*, [†] Mellier et al (1987)

The comparison between the mass distribution for the model, the red light profile given by cluster galaxies and the X-ray map leads to an interesting description of the cluster at different scales. At intermediate large scales ($1' \leq r \leq 5'$, $230 h_{50}^{-1} kpc \leq r \leq 1.15 h_{50}^{-1}$ Mpc), the three distributions look similar, with roughly the same orientation and ellipticity. This result seems to indicate that baryons actually follow the dark matter distribution, in good agreement with the SPH simulations by Navarro et al. (1995) at the same scale. There is no evidence that there is any X-ray emission at the secondary clump (infered from the lens model), but it could be below the detection limit. This heterogeneity could explain why Miralda-Escudé and Babul (1994, 1995) and Loeb & Mao (1994) have found a contradiction between the mass estimates from lensing and the X-ray temperature of the gas (assumed to be supported by thermal pressure). These authors proposed several physical mechanisms to solve this discrepancy, and one of them is the presence of bulk motions or turbulence originated by merging of substructures. An ongoing merger is indeed supported by the X-ray map and the lens modeling. Such inhomogeneity has not been observed in A370, where the X-ray peaks are centered on the two brightest galaxies (see Mellier et al. 1994). The difference could be due to a scale factor, as the redshift difference between the two clusters makes the smaller scales seen in A2218 almost undetectable in A370 (at $z = 0.374$, 1"$\equiv 6.21 h_{50}^{-1}$, the scale is therefore almost twice smaller than for A2218).

If clusters of galaxies formed hierarchically, through mergers of smaller units, a fraction of them should be seen during ongoing mergers, and there is no reason why the gas should have the same distribution as the mass. The gas experiences pressure forces and shocks which alter its trajectory with respect to the orbits followed by the merging dark matter clumps. Simulations of formation of clusters show that the mergers generally occur along filaments of the large scale structure distribution (Navarro et al. 1994, Thomas & Couchman 1993). In the case of A2218, our lensing model suggests that the mergers could take place along the line joining the cD galaxy and the galaxy #244. The dark matter is therefore elongated along this line, giving a shear that causes most of the arcs and arclets to be elongated perpendicularly to this line. The X-ray contours perpendicular to the merging line, close to the center, may be due to shocked gas, escaping to the sides, through the region of lowest pressure. The dynamics of the gas in A2218 seems to be similar to that in A370, showing a double clump and some evidence of gas moving away from a central shock.

*Acknowledgements.* We thank B. Fort, G. Mathez, R. Ellis, G. Mamon, J. Bartlett and A. Edge for interesting discussion on lensing and gas dynamics, and G. Soucail for her comments on high-redshift sources. We are very grateful to Dr. G. Bruzual for allowing the use of his code for the spectral evolution of galaxies. This work is part of the Toulouse-Barcelona ESO/CFHT - keyprogram "Arc survey". Part of this work was supported by the French Centre National de la Recherche Scientifique, by the French Groupe de Recherche Cosmologie, the EC HCM Network CHRX-CT92-0044 and by the Spanish DGICYT program PB90-0448 of the Ministerio de Educación y Ciencia. JPK acknowledges support via EC fellowship. JM is grateful for financial support from the W. M. Keck Foundation, and thanks Toulouse University and the University of Barcelona for their hospitality during a visit in which part of this work was done.

J.P. Kneib et al.: Dynamics of Abell 2218   15